\documentstyle[12pt]{article}

\newcommand{\ben}{\begin{enumerate}}
\newcommand{\een}{\end{enumerate}}
\newcommand{\be}{\begin{equation}}
\newcommand{\ee}{\end{equation}}
\newcommand{\bse}{\begin{subequation}}
\newcommand{\ese}{\end{subequation}}
\newcommand{\bea}{\begin{eqnarray}}
\newcommand{\eea}{\end{eqnarray}}
\newcommand{\bc}{\begin{center}}
\newcommand{\ec}{\end{center}}

\newcommand{\brq}[2]{\langle #1|#2\rangle}
\newcommand{\bra}[1]{\langle #1|}
\newcommand{\ket}[1]{|#1\rangle}
\newcommand{\lp}{\left(}
\newcommand{\rp}{\right)}

\newcommand{\lc}{\left[}
\newcommand{\rc}{\right]}
\newcommand{\lll}{\left\{}
\newcommand{\ry}{\right\}}

\newcommand{\uno}{1\!\mbox{l}}

\begin{document}
\begin{titlepage}
\hfill{Preprint {\bf SB/F/99-266}}
\hrule \vskip 2.5cm
\centerline{\bf SPIN OBSERVABLES AND PATH INTEGRALS}
\vskip 2cm
\centerline{J.A. L\'opez and J. Stephany \footnote{Regular
Associate of the ICTP}}
\vskip 4mm
\centerline{\it Universidad
Sim\'on Bol\'{\i}var, Departamento de F\'{\i}sica,}
\centerline{\it Apartado Postal 89000, Caracas 1080-A, Venezuela.}
\centerline{\it e-mail: stephany@usb.ve, murdock@fis.usb.ve}
\vskip 1cm

{\bf Abstract} \vskip 3mm \noindent We discuss the formulation of
spin observables associated to a non-relativistic spinning
particles in terms of grassmanian differential operators. We use
as configuration space variables for the pseudo-classical
description of this system the positions $x$ and a Grassmanian
vector $\vec\epsilon$. We consider an explicit discretization
procedure to obtain the quantum amplitudes as path integrals in
this superspace. We compute the quantum action necessary for this
description including an explicit expression for the boundary
terms. Finally we shown how for simple examples, the path integral
may be performed in the semi-classical approximation, leading to
the correct quantum propagator. \vskip 2cm \hrule
\bigskip
\centerline{\bf UNIVERSIDAD SIMON BOLIVAR}
\vfill
\end{titlepage}

\section{Introduction}
The path integral \cite{pi} formulation of the quantum mechanics
of fermionic systems is usually associated with the introduction
of Grassmann variables \cite{fpi} as pseudoclassical configuration
variables (Nevertheless see Ref. \cite{barut} for an alternative
approach in the relativistic case). The path integral treatment of
a single fermionic degree of freedom is very well understood but,
surprisingly, the extension of the formalism to a space time
description of relativistic and non relativistic spinning
particles or to the solution of potential problems has not been
developed yet. Several approaches \cite{fpi} using different sets
of Grassmannian non-commuting variables have been proposed in the
literature allowing to construct a pseudo-classical description of
the dynamics of the spinning particle, (or pseudomechanics
\cite{nrc}) for the non-relativistic case, and for the Dirac
electron \cite{rc} but this is not enough for a quantum
description. With the pseudo-classical action identified, in order
to write down a path integral expression for the fermionic
propagator, it is necessary to construct an explicit
representation of the spin observables and the polarized states of
the particle in the Hilbert space associated to the Grassmann
variables. This will we done in what follows. On the other hand,
in the usual approach, \cite{nrc} \cite{rc} this obstacle is
bypassed by showing, instead that the constraint which emerge from
imposing a variational principle to the action functional is
equivalent in the operatorial formalism to the wave equation.Then
the form of the propagator is borrowed from this formalism. This
strategy, although enlightening from the conceptual point of view
is not useful for computational purposes. Another path integral
formalism, which is based on the use of Grassmannian coherent
states \cite{cs}, has been devised for the description of
fermionic systems. It allows the computation of the propagator and
bound state energies but the relation of this formalism with the
pseudo-classical description is not completely clear and in
particular does not provide a direct interpretation of the path
integral as a sum over histories in configuration space. In what
follows we also discuss how we can get this interpretation for
spinning particles. First, we show that with an explicit
realization of the spin observables one can represent the spin
polarized states in the Grassmannian sector of the superspace.
Then we derive the path integral formulation of the
non-relativistic electron as a sum over histories directly from
the pseudo-classical description.
 Finally, we show that being careful with the
boundary conditions of the Grassmann functions one is able to
compute the probability amplitudes using a semiclassical
expansion.

\section{Wave functions and spin observables}

Consider a real Grassmannian vector $\vec{\epsilon}$ satisfying
the anti-conmutation relations, $$
\epsilon_i\epsilon_j=-\epsilon_j\epsilon_i $$ and the
super-configuration space of coordinates
($\vec{x}$,$\vec{\epsilon}$). Let us consider wave functions of
both $\vec{x}$ and $\vec{\epsilon}$ with the general expansion ,
\be
\phi(x,\vec\epsilon)=\phi(x)+\phi_i(x)\epsilon_i+
\phi_{ij}(x)\epsilon_i\epsilon_j+
\phi_{ijk}(x)\epsilon_i\epsilon_j\epsilon_k. \ee

The functions $\phi_{ij}(x)$ and $\phi_{ijk}$ are anti-symmetric.
The wave functions belong to a $8$-dimensional complex vectorial
space, $(C\!\!\!\!/\,^{8})$. The internal product in this space
may be defined by,
\be
\brq{\phi_1}{\phi_2}=\int{}d\epsilon_3d\epsilon_2d\epsilon_1(\phi^\dagger)I_\epsilon\phi_2.
\label{pin} \ee where $d\epsilon_3d\epsilon_2d\epsilon_1$ is the
Berezin integration measure. The operator $I_\epsilon$ is,
\be
I_\epsilon=\frac{\epsilon_{ijk}}{3!}\lp\epsilon_i+
\partial_i\rp\lp\epsilon_j+\partial_j\rp\lp\epsilon_k+\partial_k\rp.
\ee and $\partial_m=\frac{\partial}{\partial\epsilon_m}$ are the
Grassmannian right derivatives which satisfy,
$
\partial_m\epsilon_k=\delta_{mk}$ and $\mbox{
}\partial_m\epsilon_k\epsilon_j=\delta_{mk}\epsilon_j-\delta_{mj}\epsilon_k.
$
If one considers the eight independent functions in $\phi$ as the
components of a vector in $(C\!\!\!\!/\,^{8})$, the internal
product defined in (\ref{pin}) corresponds to the usual product in
$(C\!\!\!\!/\,^{8})$. The $\delta$ function in the odd sector is
given by
\begin{equation}
\delta(\vec\epsilon'-\vec\epsilon)=\lp\epsilon_1'-\epsilon_1\rp\lp\epsilon_2'-
\epsilon_2\rp\lp\epsilon_3'-\epsilon_3\rp{}.
\end{equation}
Introduce the non-Hermitian position operator $E$ in the odd
sector of the configuration space and the continuous set of
eigenvectors $\ket{\vec{\epsilon}}$ of $E$. Similarly as in the
coherent state representation we have the relations
\begin{equation}
\brq{\vec\epsilon}{\vec\epsilon'}=e^{\vec\epsilon'.\vec\epsilon}.
\end{equation}
An arbitrary wave function $\phi(x,\epsilon_{i})$ is represented
in Dirac notation in the form,
$
\phi(x,\epsilon_{i})=\brq{\vec{x},\vec\epsilon}{\phi} $. The
identity operator in the odd sector is
\begin{equation}
\uno=\int{}d\epsilon_{3}d\epsilon_{2}d\epsilon_{1}\ket{\vec\epsilon'}I_{\epsilon}
\bra{\vec\epsilon},
\end{equation}
and we note also that,
$
I_\epsilon{}e^{\epsilon'\epsilon}=\delta(\epsilon-\epsilon') $.
The physical sector of this space should be expanded by the spin
polarized states. To represent the spin observables $\vec{S}$ we
introduce the differential operators,
\be
\label{spin}
S_i=-\frac{i}{4}\epsilon_{ijk}\lp\epsilon_j+\partial_j\rp\lp\epsilon_k+\partial_k\rp
\ee which satisfy the angular momentum algebra,
$[S_i,S_j]=i\epsilon_{ijk}S_k$. This representation of the spin
observables is the natural generalization of the usual
representation of the fermionic position-momentum algebra which
led to the coherent state formulation of the path
integral\cite{bt}. It has been discussed also in Ref.
\cite{mankoc}. Looking at things from another point of view, the
Hilbert space of states $\ket{\vec\epsilon}$ and the operators
(\ref{spin}) provide a fermionic coherent state representation of
the $SU(2)$ algebra alternative to the bosonic approach
\cite{hioe}. This construction may be generalized to other groups.
The complete set of eigenfunctions of $S_3$ is given in the
following table.
\begin{center}
\begin{tabular}{|c|c|c|}\hline
$f_{\lambda}^n$&$\lambda$&$\phi(\vec\epsilon)$\\\hline
$f_{+}^1$&$\frac{1}{2}$&$1-i\epsilon_1\epsilon_2$\\\hline
$f_{+}^2$&$\frac{1}{2}$&$\epsilon_3-i\epsilon_1\epsilon_2\epsilon_3$\\\hline
$f_{+}^3$&$\frac{1}{2}$&$\epsilon_1+i\epsilon_2$\\\hline
$f_{+}^4$&$\frac{1}{2}$&$-\epsilon_1\epsilon_3-i\epsilon_2\epsilon_3$\\\hline

$f_{-}^4$&$-\frac{1}{2}$&$1+i\epsilon_1\epsilon_2$\\\hline
$f_{-}^3$&$-\frac{1}{2}$&$\epsilon_3+i\epsilon_1\epsilon_2\epsilon_3$\\\hline
$f_{-}^2$&$-\frac{1}{2}$&$\epsilon_1-i\epsilon_2$\\\hline
$f_{-}^1$&$-\frac{1}{2}$&$-\epsilon_1\epsilon_3+i\epsilon_2\epsilon_3$\\\hline

\end{tabular}
\end{center}
The eigenvalues of $S_3$, denoted by $\lambda$ are which, as can
be seen, degenerate. To construct a particular base of states, we
take a linear combination of them in such a way that the action of
the up and down operators $S_+=S_1+iS_2$ and $S_-=S_1-iS_2$ is
well defined. A possible choice of the eigenfunctions which
represent the polarized states is
\begin{eqnarray}
\brq{\vec\epsilon}{+}&=&(1+\epsilon_3)(1-i\epsilon_1\epsilon_2),\\
\nonumber
\brq{\vec\epsilon}{-}&=&(1-\epsilon_3)(\epsilon_1-i\epsilon_2).
\end{eqnarray}

\section{The action functional and the path integral}
Let us consider now a spinning particle whose dynamics is
determined by the Hamiltonian operator $H$. We want to discuss a
discretization procedure in the trajectories of the particle in
the configuration space which allows to represent the Green
functions in terms of a path integral. The evolution operator is
given by,
\begin{equation}
U(t_f-t_i)=e^{-iH(t_f-t_i)}.
\end{equation}
with matrix elements,
$
U(x,\vec\epsilon,t;x',\vec\epsilon',t')=\bra{x',\vec\epsilon'}U(t,t')\ket{x,\vec\epsilon}.
$
For the spin polarized states ($k=+,-$), we define the physical
propagator,
$
K(k,t_f;j,t_i=\bra{x_f,k}U\ket{x_i,j}$ which may be projected in
the form,
\be
\label{prop}
 K(k,t_f;j,t_i)
=\int{}d\epsilon_i\int{}d\epsilon_f\brq{k}{\epsilon_f}
I_{\epsilon_f}\bra{\epsilon_f}U(x_f,\epsilon_f,t_f;x_i,\epsilon_i,t_i)\ket{\epsilon_i}
I_{\epsilon_i}\brq{\epsilon_i}{j}. \ee (we drop the arrow on the
Grassmann coordinates).Now consider a discretization
$\{t_{1},t_{2},\ldots.,t_{2N}\}$ with $\delta=t_{k}-t_{k-1}$ of
the time interval, and using the resolution of unity let us
compute the matrix element of the evolution operator. We get,
\begin{eqnarray*}
&&\bra{x_f,\epsilon_f}e^{-iH(t_f-t_i)}\ket{x_i,\epsilon_i}=\\
\nonumber
&&\lim_{{2N}\rightarrow\infty}\int{}\Pi_{k=1}^\infty\frac{dp_k}{2\pi}\int{}\Pi_{j=1}^\infty{}
dx_j\int{}d\epsilon_1'\int{}d\epsilon_1\int{}d\epsilon_2'\int{}d\epsilon_2...
\int{}d\epsilon_{2N}' \int{}d\epsilon_{2N}\\ \nonumber
&&\bra{x_f,\epsilon_f}\ket{p_{2N},\epsilon_{2N}}I_{\epsilon_{2N}}\bra{p_{2N},
\epsilon_{2N}}e^{-iH\delta}\ket{x_{2N},\epsilon_{2N}'}I_{\epsilon_{2N}'}\bra{x_{2N},
\epsilon_{2N}'}...\\ \nonumber
&&...\bra{\epsilon_k,p_k}e^{-iH\delta}\ket{x_k,\epsilon_k'}I_{\epsilon_k'}
\bra{x_k,\epsilon_k'}\ket{p_{k-1},\epsilon_{k-1}}I_{\epsilon_{k-1}}\bra{p_{k-1},\epsilon_{k-1}}...\\
\nonumber
&&...\bra{\epsilon_1,p_1}e^{-iH\delta}\ket{x_1,\epsilon_1'}I_{\epsilon_1'}
\brq{x_1,\epsilon_1'}{x_i,\epsilon_i}.
\end{eqnarray*}

The general term can be expanded in the form,
\begin{eqnarray*}
&&I_{\epsilon_k}\bra{\epsilon_k,p_k}e^{-iH\delta}\ket{x_k,\epsilon'_k}I_{\epsilon_k'}
\brq{x_k,\epsilon_k'}{p_{k-1},\epsilon_{k-1}}= \\ \nonumber
&&I_{\epsilon_k'}e^{-iH(p_k,x_k,\epsilon_k)\delta}e^{-ip_kx_k}e^{\epsilon_k'\epsilon_k}
I_{\epsilon_k'}e^{ip_{k-1}x_{k-1}}e^{\epsilon_{k-1}\epsilon_k'}=
\\ \nonumber
&&e^{-iH_0(p_k,x_k)\delta}e^{-ip_kx_k}e^{ip_{k-1}x_{k-1}}I_{\epsilon_k'}
e^{-iH_1(x_k,\epsilon_k)\delta}
e^{\epsilon_k'\epsilon_k}I_{\epsilon_k'}e^{\epsilon_{k-1}\epsilon_k'}
\end{eqnarray*}
were we use that the hamiltonian function satisfies,
\be
H(p_k,x_k,\epsilon_k)=H_0(p_k,x_k)+H_1(x_k,\epsilon_k). \ee This
is obvious if $H$ does not depend on the $\epsilon$ derivatives,
but it is also true in the general case due to the external
integrals. Let us focus in the Grassmann sector alone and note
that under the integral sign we have,
\begin{eqnarray*}
&&\brq{\epsilon_f}{\epsilon_{2N}}I_{\epsilon_{2N}}
e^{-iH_1(x_{2N},\epsilon_{2N)}\delta}
e^{\epsilon_{2N}'\epsilon_{2N}}I_{\epsilon_{2N}'}e^{\epsilon_{{2N}-1}\epsilon_{2N}'}...
I_{\epsilon_1}e^{-iH_1(x_1,\epsilon_1)\delta}e^{\epsilon_i\epsilon_1}=
\\ \nonumber
&&e^{{\frac{1}{2}}\epsilon_{2N}\epsilon_f}e^{-iH(p_{2N},x_{2N},\epsilon_{2N})\delta+{\frac{1}{2}}\epsilon_{2N}
\lp\epsilon_f-\epsilon_{{2N}-1}\rp}e^{-iH(p_{2N-1},x_{2N-1},\epsilon_{2N-1})\delta+
{\frac{1}{2}\epsilon_{{2N}-1}}\lp
\epsilon_{2N}-\epsilon_{{2N}-2}\rp}...\\ \nonumber
&&e^{-iH(p_{1},x_{1},\epsilon_{1})\delta+\frac{1}{2}\epsilon_1\lp
\epsilon_2-\epsilon_i\rp}e^{{\frac{1}{2}}\epsilon_i\epsilon_1}.
\end{eqnarray*}

The terms at the end of the interval are of the form,
\begin{eqnarray}
\vec\epsilon_{2N}\vec\epsilon_f&\approx&(\vec\epsilon_f-\delta\dot{\vec\epsilon_f})\vec\epsilon_f=
\delta\vec\epsilon_f\dot{\vec\epsilon_f},\\
\vec\epsilon_i\vec\epsilon_1&\approx&\vec\epsilon_i(\vec\epsilon_i+\delta\dot{\vec\epsilon_i})=
\delta\vec\epsilon_i\dot{\vec\epsilon_i}.
\end{eqnarray}
In the limit ${2N}\rightarrow\infty$ ($\delta\rightarrow{}0$),
they reduce to boundary terms which depend only of initial and
final values
\begin{equation}
g(\vec\epsilon_i,\vec\epsilon_f)=\lim_{\delta\rightarrow{}0}\frac{1}{2}\lll
\int_{t_i}^{t_{i}+\delta}\vec\epsilon\dot{\vec\epsilon}dt+\int_{t_{f}-\delta}^{t_f}
\vec\epsilon\dot{\vec\epsilon}dt\ry. \label{borde}
\end{equation}
Incorporating the bosonic sector we are left with,
\begin{equation}
\label{IF}
U(\vec\epsilon_f,t_f;\vec\epsilon_i,t_i)=\int{}D[\epsilon]D[x]D[p]e^{g(\vec\epsilon_i,\vec\epsilon_f)+i\int\lll{}\dot{\vec{x}}\vec{p}-\frac{i\vec\epsilon\dot{\vec\epsilon}}{2}-H(x,p,\epsilon)\ry{}dt}.
\end{equation}
The action functional recovered in the measure of the path
integral is the one that appears in the pseudoclassical
description of the spinning particle \cite{nrc}. To compute the
quantum ampitude between physical states, one introduces
(\ref{IF}) in (\ref{prop}).

\section{The semiclassical approximation and the variational principle}

Bosonic path integrals with quadratic potentials may be computed
using a semiclassical approximation \cite{pi}. In this section we
show that a similar result holds also in the case under
consideration if proper care is given to the boundary terms. The
point here is that, since the equations for $\vec\epsilon$ are
first order, it is not possible in general to find trajectories
$x(t)$ and $\vec\epsilon(t)$, extremals of $S$ in the time
interval $t_f -t_i$, with $x(t_i)=x_i$, $x(t_f)=x_f$,
$\epsilon(t_i)=\epsilon_i$ and $\epsilon(t_f)=\epsilon_f$. So we
introduce two Lagrange multipliers $\pi_i,\pi_f$ and consider
instead an extended action
\begin{equation}
S^*[\epsilon(t),\pi_i,\pi_f]=S[\epsilon(t)]+\pi_i(\epsilon(t_i)-\epsilon_i)-\pi_f(\epsilon(t_f)-\epsilon_f).
\end{equation}
The equations of motion are
\begin{eqnarray}
2\lp{}\lp{}\frac{\partial{}L}{\partial{}\dot{\epsilon}}\rp{}(t_f)-\pi_f\rp{}\delta(t-t_f)-
2\lp{}\lp{}\frac{\partial{}L}{\partial{}\dot{\epsilon}}\rp{}(t_i)-\pi_i\rp{}\delta(t-t_i)=0,
\\\nonumber x(t_i)=x_i, x(t_f)=x_f. \label{eq_lag}
\end{eqnarray}
Now we can fix the values of the Lagrange multipliers to guarantee
that the boundary conditions, which here appear as independent
equations, are satisfied. In fact, we still have the freedom to
fix $\pi_i$ to zero. Then the solution to the equations may be
written in the form
\be
\vec{\epsilon}_{class}=\vec{\epsilon_{0}}+\vec{\delta_{\epsilon}}\Theta(t-t_{f})
\ee where $\vec{\epsilon_{0}}$ satisfies
\begin{equation}
\dot{\vec\epsilon_{0}}=i\frac{\partial{}H}{\partial\vec\epsilon_{0}}.
\label{eq.clas}
\end{equation}
and $\delta_{\epsilon}$ is a jump at the end of the trajectory. To
perform the semiclassical expansion let us consider first the free
case with the action given by,
\begin{equation}
iS=g(\vec\epsilon_i,\vec\epsilon_f)+i\int\lc{}-
\frac{i\vec\epsilon\dot{\vec\epsilon}}{2}\rc{}dt. \label{ac_impar}
\end{equation}
The solution to the equations of motion which satisfies the
boundary condition is simply
\begin{equation}
\label{free}
\vec\epsilon(t)=\vec\epsilon_i+2(\vec\epsilon_f-\vec\epsilon_i)\Theta(t-t_f).
\end{equation}
Consider the path integral (\ref{IF}) computed in the previous
section with the boundary term (\ref{borde}), and let us perform
an expansion around $\vec{\epsilon}_{class}$,
\par
\begin{equation}
\vec\epsilon(t)=\vec\epsilon_{class}(t)+\vec\xi(t). \label{expan}
\end{equation}
Substituting (\ref{expan}) and (\ref{free}) in (\ref{IF}) we get
the expected result,
\begin{equation}
U(\vec\epsilon_f,t_f;\vec\epsilon_i,t_i)=Ne^{\epsilon_i\epsilon_f}.
\end{equation}
Let us turn out our attention to a more general case. The most
general even Hamiltonian function has the form,
\be
H(x,p,\epsilon)=H_0(x,p)+ H_{ij}(x)\epsilon_i\epsilon_j. \ee Then,
the equation of motion is linear in $\epsilon$. Using the equation
of motion for $\epsilon_{0}$, and the linearity of the equation of
motion it is readily seen that the boundary term takes the form,
\begin{equation}
g(\vec\epsilon_i,\vec\epsilon_f)=\frac{\vec\epsilon_0(t_f)\vec\epsilon_f}{2}.
\label{borde.def}
\end{equation}
This result generalizes for the interacting case the expression
obtained by Galvao and Teitelboim \cite{nrc}. Consider again
expressions of the form (\ref{expan}) and (\ref{expan}).
Substitution in (\ref{IF}) leads us to the expression,
\begin{eqnarray}
g(\vec{\epsilon_i},\vec{\epsilon_f}) +
iS=\vec{\epsilon_0}(t_f)\vec{\epsilon_f}+\\\nonumber
i\int_{t_i}^{t_f}dt{}\lll{}\lc{}-i\frac{\vec{\epsilon_0}
\dot{\vec{\epsilon_0}}}{2}+
H_{0}(x,p)+H_{ij}\epsilon_{0i}\epsilon_{0j}\rc{}+\lc{}-
\frac{i}{2}\vec{\xi}\dot{\vec{\xi}}+H_{ij}\xi_{i}\xi_{j}\rc\ldots\ry{}
\end{eqnarray}
for the exponent. (The dots appear to denote possible bosonic
contributions). Then, in the case when the spin degrees of freedom
factorize, we get the following simple expression for the matrix
elements
\begin{equation}
U(\vec\epsilon_f,t_f;\vec\epsilon_i,t_i)=Ne^{2g(\vec\epsilon_f,\vec\epsilon_i)}.
\end{equation}
where $N$ is a normalizacion constant.
\section{Spin precession}
Let us recover the known results for particle in a uniform
magnetic field for example. The action is given by
\begin{equation}
S=g(\vec\epsilon_i,\vec\epsilon_f)+\int\lp{}\dot{\vec{x}}\vec{p}-
\frac{i}{2}\vec\epsilon\dot{\vec{\epsilon}}+\frac{p^2}{2m}+\frac{q}{2m}
\lp\frac{\epsilon_{ijk}}{2}i\epsilon_j\epsilon_k\rp{}B_i\rp{}dt.
\end{equation}
Defining,
\begin{equation}
M_{jk}=\frac{q}{2m}\frac{\epsilon_{ijk}}{2}B_i,
\end{equation}
we are left with the Lagrangian,
\begin{equation}
L=-\frac{i\vec\epsilon\dot{\vec\epsilon}}{2}+i\frac{\vec\epsilon^T{}M\vec\epsilon}{2}.
\end{equation}
The equation of motion is simply,
$
\dot{\vec\epsilon}_{0}(t)=M\vec\epsilon_{0}(t) $, and the
classical trajectory with arbitrary boundary conditions is given
by,
\begin{eqnarray}
\vec\epsilon_{class}(t)&=&e^{M(t-t_i)}\vec\epsilon_i+\vec\delta_\epsilon{}\Theta(t-t_f),\\
\nonumber
\dot{\vec\epsilon}_{class}(t)&=&Me^{M(t-t_i)}\vec\epsilon_i+\vec\delta_\epsilon\delta(t-t_f),
\end{eqnarray}
Defining $\omega=\frac{qB}{m}$ we have,
\begin{equation}
e^{M(t-t_i)}=\pmatrix{cos(\omega{}(t-t_i))&sen(\omega{}(t-t_i))&0\cr
-sen(\omega{}(t-t_i))&cos(\omega{}(t-t_i))&0\cr 0&0&1\cr}.
\end{equation}
In this case the boundary term $g(\vec\epsilon_f,\vec\epsilon_i)$
is nontrivial and takes the form
 \begin{equation}
g(\vec\epsilon_f,\vec\epsilon_i)=\frac{\vec\epsilon_i^te^{-M(t_f-t_i)}\vec\epsilon_f}{2}.
\end{equation}
According with the discussion of the previous section the we have
now,
\begin{equation}
U(\vec\epsilon_f,t_f;\vec\epsilon_i,t_i)=e^{\vec\epsilon_i^te^{-M(t_f-t_i)}\vec\epsilon_f}.
\end{equation}
To recover the standard result we compute the time evolution of an
arbitrary wave function
\begin{equation}
\phi(\vec\epsilon_f,t_f)=\int{}d\epsilon_iI_{\vec\epsilon_f}
e^{\vec\epsilon_i^te^{-M(t_f-t_i)}\vec\epsilon_f}\phi(\vec\epsilon_i,t_i)=
\phi(e^{-M(t_f-t_i)}\vec\epsilon_f,t_i).
\end{equation}
With the initial state,
$\ket{\phi_i}=cos(\frac{\theta}{2})e^{\frac{-i\varphi}{2}}\ket{+}+sen(\frac{\theta}{2})
e^{\frac{i\varphi}{2}}\ket{-}$ and $\vec{B}$ directed in the
$x_{3}$ direction, we get
\begin{eqnarray}
\bra{\phi(t)}S_3\ket{\phi(t)}&=&cos(\theta),\\ \nonumber
\bra{\phi(t)}S_1\ket{\phi(t)}&=&sen(\theta)cos(\varphi+\omega{}t),\\
\nonumber
\bra{\phi(t)}S_2\ket{\phi(t)}&=&sen(\theta)sen(\varphi+\omega{}t).
\end{eqnarray}

\section{Conclusion }
In this paper we have assembled many sparse elements of the the
theory of spinning particle already found in the literature, and
developed a little some of them, to construct a path integral
representation of the quantum amplitudes of a non-relativistic
electron in an external electromagnetic field. This fermionic path
integral shares the interpretation of a sum over (pseudo)
classical histories with its bosonic counterpart. The clue in this
approach is to build up the path integral from the explicit
realization of the spin operators. The main technical point in the
computations concerns the correct handling of the boundary
contributions. There are various natural ways to develop further
the work presented in this paper. First, one can extend the
computational techniques to cases where the spin and the
translational degrees of freedom are mixed by the interaction (For
example in Ref.\cite{calvo}). One can also generalize this
approach to the relativistic Dirac particle as we discuss
elsewhere \cite{lopez}. Finally the relation between the
Grassmannian representation of the spin observables and the
fermionic $SU(2)$ coherent states may be generalized for other
groups.

\section{Acknowledgments}
It is a pleasure to thank the Organizing Committee and all the
people involved in the organization of the VI Wigner Symposium for
the high level scientific atmosphere and pleasant environment that
they created even in adverse conditions. J.S wish also thank
J.D.Vergara for very useful discussions on the topics treated in
this work

\end{document}